\documentclass[conference]{IEEEtran}
\usepackage{cite}
\usepackage{amsmath,amssymb,amsfonts}
\usepackage{algorithmic}
\usepackage{graphicx}
\usepackage{caption}
\usepackage{textcomp}
\usepackage{xcolor}
\usepackage{svg}
\usepackage{afterpage}
\usepackage{booktabs}
\usepackage{float}
\usepackage{subfigure}
\usepackage{subfig}
\usepackage{adjustbox}
\definecolor{darkgreen}{rgb}{0.0, 0.5, 0.0}
\usepackage{multirow}

\def\BibTeX{{\rm B\kern-.05em{\sc i\kern-.025em b}\kern-.08em
    T\kern-.1667em\lower.7ex \hbox{E}\kern-.125emX}}

\begin{document}

\title{Exploring Highly Quantised Neural Networks for Intrusion Detection in Automotive CAN}

\author{\IEEEauthorblockN{Shashwat Khandelwal \& Shanker Shreejith}
\IEEEauthorblockA{ Reconfigurable Computing Systems Lab, Electronic \& Electrical Engineering\\
Trinity College Dublin, Ireland\\
Email: \{khandels, shankers\}@tcd.ie}}



\maketitle

\begin{abstract}
Vehicles today comprise intelligent systems like connected autonomous driving and advanced driving assistance systems (ADAS) to enhance the driving experience, which is enabled through increased connectivity to infrastructure and fusion of information from different sensing modes. 
However, the rising connectivity coupled with the legacy network architecture within vehicles can be exploited for launching active and passive attacks on critical vehicle systems and directly affecting the safety of passengers. 
Machine learning-based intrusion detection models have been shown to successfully detect multiple targeted attack vectors in recent literature, whose deployments are enabled through quantised neural networks targeting low-power platforms. 
Multiple models are often required to simultaneously detect multiple attack vectors, increasing the area, (resource) cost, and energy consumption.
In this paper, we present a case for utilising custom-quantised MLP's (CQMLP) as a multi-class classification model, capable of detecting multiple attacks from the benign flow of controller area network (CAN) messages.
The specific quantisation and neural architecture are determined through a joint design space exploration, resulting in our choice of the 2-bit precision and the \text{n}-layer MLP.
Our 2-bit version is trained using Brevitas and optimised as a dataflow hardware model through the FINN toolflow from AMD/Xilinx, targeting an XCZU7EV device. 
We show that the 2-bit CQMLP model, when integrated as the IDS, can detect malicious attack messages (DoS, fuzzing, and spoofing attack) with a very high accuracy of 99.9\%, on par with the state-of-the-art methods in the literature.
Furthermore, the dataflow model can perform line rate detection at a latency of 0.11\,ms from message reception while consuming 0.23\, mJ/inference, making it ideally suited for integration with an ECU in critical CAN networks.

\end{abstract}

 \begin{IEEEkeywords}
 Controller Area Network, Intrusion Detection System,
Quantised Neural Nets, Machine Learning, FPGAs
 \end{IEEEkeywords}

\section{Introduction \& Related Works}
\begin{figure}%
    \centering
    \subfigure[\centering]{\includegraphics[width=3.95cm,height = 2.5cm]{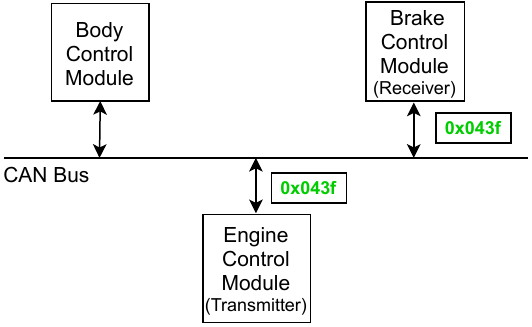} }%
    \subfigure\centering]{\includegraphics[width=3.95cm, height = 2.5cm]{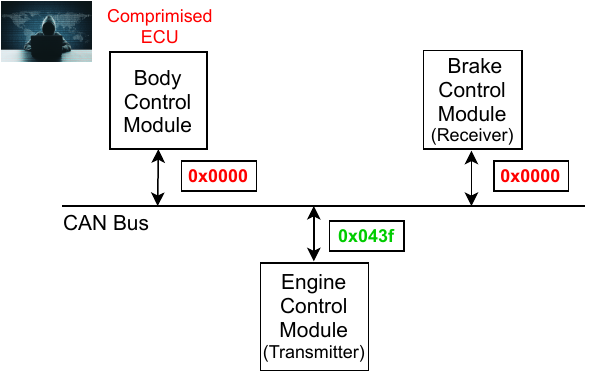} }%
    \caption{An illustration of a simple DoS attack launched through a compromised ECU. Part (a) shows the normal communication between the ECUs, while in (b), the ECU is unable to transmit messages on the CAN bus as the compromised ECU floods the bus with high priority messages.}%
    \label{fig:canbus}%
\end{figure}

Automotive networks are evolving rapidly to cater to the high-data needs for novel/intelligent capabilities that have enabled safety, infotainment, and comfort for passengers. 
These interconnected networks (via multiple network protocols) facilitate high-speed communication/information exchange between over a hundred electronic computing units (ECUs), sensors, and actuators present in modern cars. 
Among the network protocols, Controller Area Network (CAN) is used for critical communication between ECUs and it continues to be the most widely used network protocol for in-vehicle networks owing to its cost-effective nature and ease of use in control applications~\cite{CanBosch}.
Early ECUs and software functions deployed on the CAN were automatically siloed due to the limited connectivity of the vehicles then, to the external world.
Novel capabilities in vehicles today rely on connectivity to infrastructure and other vehicles to enable real-time sensing of the outside environment and also to enable unique features like remote monitoring and control of specific capabilities for diagnostics, and over-the-air upgrades. 
Researchers however have shown that such interfaces open up new avenues for injecting malicious code/messages into these previously siloed networks~\cite{nie2017free,iehira2018spoofing,cai20190}.
Figure~\ref{fig:canbus} shows an example of a simple DoS attack on a standard CAN bus.
Such attacks are largely enabled by the lack of inherent security and authentication mechanisms in CAN and similar automotive network protocols~\cite{miller2015remote,enev2016automobile}. 

To detect the onset of such intrusions on CAN, multiple intrusion detection approaches have been proposed in the literature allowing critical systems to enter into a `safe working' mode when such threats are detected.
Early intrusion detection systems (IDS) proposed were rule/specification based, which utilised a set of rules to compare known attack signatures to patterns captured from current network parameters/messages to detect unusual activity~\cite{larson2008approach,miller2013adventures,studnia2018language}. 
Recently, machine learning (ML) approaches have shown significant improvement in the detection accuracy of such threats and the ability to adapt to newer attack vectors~\cite{narayanan2015using, alshammari2018classification, yang2019tree, song2020vehicle, tariq2020cantransfer} without incurring overheads of rule-based approaches. 
In addition to native precision ML models, quantised variants of the floating-point models have shown to be viable alternatives, reducing the computational complexity, resource, and energy consumption at the expense of a slight reduction in inference accuracy. 
Quantised models are then deployed on constrained devices such as low-end FPGAs or dedicated embedded platforms~\cite{jokic2018binaryeye}.
Frameworks such as Vitis-AI and FINN can convert the Pytorch/TensorFlow representations at native precisions to quantised models and dataflow-style hardware accelerators for deployment on the FPGA~\cite{xilinxvitis,umuroglu2017finn}.
ECU architectures based on hybrid FPGAs are a promising platform for the deployment of such QNNs as IDS, while also enabling consolidation with standard ECU function(s), with clear isolation between them on the same die~\cite{khandelwal2022lightweight}. 
However, most methods so far relied on the integration of multiple accelerators, each fine-tuned to detect a subset of attacks, increasing the energy cost and area overhead.

In this paper, we explore the case for a custom quantised feed-forward network that can classify multiple attack vectors simultaneously and a dataflow-style custom quantised hardware implementation of the model.
The dedicated hardware implements IDS capability as an Advanced eXtensible Interface (AXI) slave peripheral to the ARM cores, offering complete isolation from the software tasks on them, mimicking the model ECU architecture for deploying distributed IDS in CAN networks. 
Starting from a standard multi-layer perceptron (MLP) model, we analyse and evaluate different model configurations and quantisations for weights/biases/activations to arrive at our custom model with 2-, 3- \& 4-bit precisions through a constrained design space exploration. 
Subsequently, the dataflow accelerator is generated and deployed using AMD/Xilinx's FINN framework~\cite{umuroglu2017finn}.
This integration allows the software tasks on the ECU to invoke and fully control the operation of the IDS accelerator through APIs, similar to offloads enabled by AUTOSAR abstractions. 
The key contributions of this paper are as follows: 
\begin{itemize}
    \item A feed-forward custom quantised multi-layer-perceptron based-IDS for automotive CAN achieving state-of-the-art classification accuracy across multiple attack vectors using only a single model.
    \item Exploits the tightly-coupled ECU architecture with the dataflow implementation of IDS accelerating IDS operation in full isolation.
    \item Quantify the performance and energy savings of the proposed ML model and its integration using the open CAN dataset. Our results show that the proposed IDS achieves significant improvements in terms of per-message processing latency and power consumption against the state-of-the-art IDSs in the literature.
\end{itemize}

We evaluate our approach using the openly available CAR Hacking dataset~\cite{song2020vehicle} for training and validation across multiple attack vectors captured from an actual vehicle with the entire CAN data frame used as an input feature to improve the detection performance.
We analyse the 2-, 3- \& 4-bit precision models and utilise the inference accuracy and resource utilisation parameters to arrive at the most optimal model.
Our experiments show that the proposed custom quantised (2-bit) MLP-based IDS (referred to as CQMLP-IDS) achieves an average accuracy of 99.91\% across Denial of Service (DoS), Fuzzing, and RPM-spoofing attacks among the benign flow of messages, identical to or exceeding the detection accuracy achieved by state-of-the-art \text{GPU- and CPU-based} implementations. 
The tightly integrated ECU architecture reduces the per message execution latency by 2.2$\times$ and the energy consumed by 3.9$\times$ compared to state-of-the-art IDSs proposed in the research literature. 
We also show that our (fp32) MLP model on a Jetson Nano~\cite{NanoGPUlink}, mimicking a dedicated IDS ECU, incurs 12$\times$ higher energy consumption per inference, compared to the integrated IDS-ECU.

The remainder of the paper is organised as follows: Section~\ref{sec:background} captures background information and state-of-the-art research in this area; section~\ref{sec:architecture-date} describes the proposed MLP model and design choices for the implementation; section~\ref{sec:experiments-date} outlines the experiment setup and results; and we conclude the paper in section~\ref{sec:conclusion}.
\section{Background and Related works}\label{sec:background}
\subsection{Controller Area Network}
In-vehicle networks enable distributed ECUs to exchange control and data messages to achieve the global functions of the vehicle. 
Multiple protocols are used in vehicular systems to cater to different functions based on their criticality and to optimise the cost of E/E systems. 
CAN~\cite{CanBosch} and CAN-FD~\cite{hartwich2012can} continue to be the most widely used protocol today due to their lower cost, flexibility, and robustness. 
The broadcast CAN bus uses carrier sense multiple access with the collision avoidance and arbitration priority (CSMA/CA-AP) access/arbitration mechanism to control access to the bus using the CAN ID allocated to each message.
This enables higher priority messages to be processed first hence efficiently handling messages from safety-critical applications.
CAN also supports multiple data rates (125\,Kbps to 1\,Mbps) and multiple modes of operation (1-wire, 2-wire) to cater to a range of critical and non-critical functions in vehicles. 
Despite this robustness, CAN is inherently insecure: there is no built-in mechanism (like message encryption) in the network to authenticate the transmitter, receiver, or the message content itself~\cite{8658720}. 
This makes CAN vulnerable to simple yet efficient active message injection attacks like fuzzing, spoofing, replay attacks, and Denial of Service (DoS) attacks~\cite{mukherjee2016practical,enev2016automobile,koscher2010experimental,palanca2017stealth}.
Researchers have explored different flavours of IDSs from rule-based approaches to machine learning-based methods to address CAN's existing vulnerabilities.
Rule-based approaches are further classified into flow-based and payload-based. 
While flow-based approaches identify traits like message frequency and/or interval for the network to detect abnormalities~\cite{vuong2015performance}, payload-based approaches use the data segment in CAN frames to detect abnormal sequences of instructions and or data~\cite{stabili2017detecting}.
Purely payload-based methods, however, fail to capture the timing/sequence of messages in an attack, leading to parts of attack messages being flagged as benign. 
Hybrid schemes use both the message timing/frequency and the payload information to capture a more holistic view of the network, allowing them to extract specific signatures of transmitting ECUs, receiving ECUs, and messages~\cite{weber2018embedded}. 
Machine learning-based techniques further generalises such techniques when trained with large datasets through classification approaches, sequential techniques, and deep learning-based schemes.

\subsection{Machine Learning based IDSs}
ML-based IDSs discussed in the literature mostly use a combination of input features, as shown in table~\ref{table:inpfeatures}. 
In~\cite{song2020vehicle}, the authors propose a reduced inception net architecture as an IDS using deep convolutional neural networks. 
They achieve over 99\% accuracy for DoS, fuzzing \& spoofing attacks.
In~\cite{10105249}, authors present a graph based intrusion detection protocol and propose improvements over existing graph based IDS techniques.
In~\cite{10141599}, a transformer network-based IDS is proposed which demonstrates very high classification accuracy at the cost of higher detection latency. 
In~\cite{seo2018gids}, the authors propose a GAN-based IDS and achieve an average accuracy of 97.5\% for the same attacks.
More complex ML architectures like temporal convolution with global attention~\cite{cheng2022tcan}, a combination of convolutional neural networks (CNN) and long short-term memory (LSTM) cells using both supervised and unsupervised approaches~\cite{agrawal2022novelads,lo2022hybrid}, gated recurrent units (GRU) networks~\cite{ma2022gru} have been shown to improve detection accuracy.
In~\cite{de2021efficient}, the authors use an iForest anomaly detection algorithm as an intrusion prevention system (IPS) to detect fuzzing and spoofing (RPM \& Gear) attacks and mark the message as an error preventing its propagation to other ECUs; however, this can cause multiple messages to be dropped from the bus in case of false positives or DoS attacks.
The key challenge of ML-based approaches is their deployment as an in-vehicle ECU. 
Most approaches rely on high-performance GPUs to meet the inference deadlines~\cite{seo2018gids,song2020vehicle,desta2020mlids,ma2022gru,agrawal2022novelads,cheng2022tcan}, while others rely on dedicating full ECUs for IDS~\cite{yang2021mth,de2021efficient}; both approaches incur additional overheads in energy budget and weight, making them less suited for distributing IDS among different network segments. 
Similarly, almost all methods presented in the literature require multiple uniquely trained models to detect all possible threat vectors, which incur much higher resources and energy than a singular model which can classify multiple threat vectors.

\begin{table}[t]
\centering
\caption{Input features used by the IDSs \& IPSs proposed in the research literature.}
\begin{tabular}{ll}
\toprule
\textbf{Models} & \textbf{Input Features Used}                 \\
\midrule
GIDS~\cite{seo2018gids}            & CAN ID                                       \\
DCNN~\cite{song2020vehicle}            & CAN ID                                       \\
Rec-CNN~\cite{desta2022rec}         & CAN ID                   \\
G-IDCS~\cite{10105249}                 & CAN ID \\
TAN-IDS~\cite{10141599}             & CAN ID \\
iForest~\cite{de2021efficient}         & Data Field                                   \\
MLIDS~\cite{desta2020mlids}           & CAN ID + Data Field                          \\
NovelADS~\cite{agrawal2022novelads}        & CAN ID + Data Field                          \\
TCAN-IDS~\cite{cheng2022tcan}        & CAN ID + Data Field                          \\
MTH-IDS~\cite{yang2021mth}         & CAN ID + Data Field \\
HyDL-IDS~\cite{lo2022hybrid}           & CAN ID + Data Field + DLC                         \\
GRU~\cite{ma2022gru}               &  CAN ID + Data Field + DLC \\ 
\textbf{CQMLP-IDS (proposed)}         & CAN ID + Data Field \\
\bottomrule
\end{tabular}
\label{table:inpfeatures}
\end{table}

\section{System architecture}\label{sec:architecture-date}

\begin{figure}[t!]
    \centering
    \includegraphics[scale = 0.65]{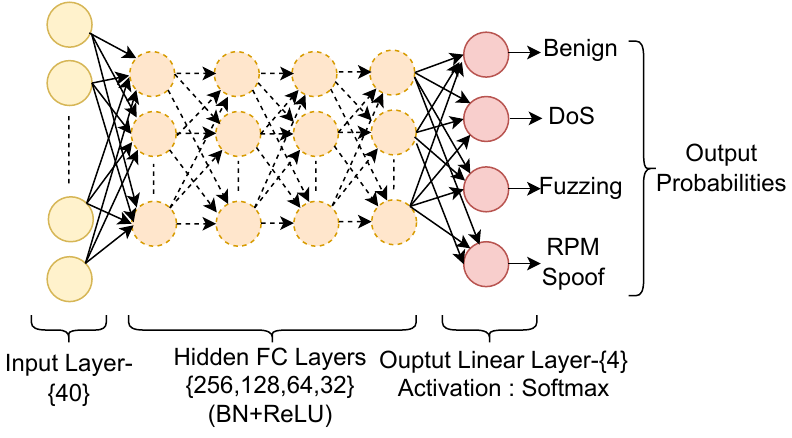}
    \caption{The proposed MLP model as a multi-attack detection IDS.}
    \label{fig:model}
    \vspace{-2mm}
\end{figure}

\subsection{MLP model for IDS}
To determine the best ML model for IDS, we profiled different ML architectures to find a model with high classification accuracy which also offers low computational complexity.
We observed that while CNNs were effective for classification when using only CAN IDs, MLPs provided more accurate detection (accuracy, false positives, and false negatives) when using the entire message contents at much lower computational complexity; since message data can contain malicious content, we chose MLPs as our choice of architecture and entire message as the input feature~\cite{9912045,10035170}. 
A concatenation of \textit{n = 4} successive messages (CAN IDs + payload) is chosen as the input feature based on this testing.
Note that during training/testing, if the block has even one attack message, the entire block is treated as an attack block.
In a real system, this means that a window of 4 messages containing a single attack message will be detected as a malicious window, allowing a sequence of events leading to and after the attack message as potential threats. 
Though conservative, this approach enables the model to better capture normal operating sequences during training and provides a better balance to the dataset.  
We observed that using a block with a higher number of messages ( \textit{$>$ 4} ) improves the classification accuracy as the model has more input data to make a decision; however, grouping more messages also affects the detection latency from the arrival of the \textit{‘infected’} message, leading to our choice of \textit{n = 4}.

Our final MLP model consists of 5 \emph{Linear} layers implemented with the 256, 128, 64, 32 \& 4 units at each layer as shown in Figure~\ref{fig:model}. 
The time-series data from the input feature buffer is fed as input to the input layer with 40 units. 
Subsequent layer(s) operate on the output of the previous layer, with increasing complexity to perform classification.
Batch Normalisation was used between the \emph{linear} layers to prevent over-fitting and to improve the learning efficiency during the training phase. 
The output \emph{linear} layer uses a \emph{softmax} activation at the output to estimate the probability of the message being benign, DoS, fuzzing, or RPM-spoofed.
The model is defined in Pytorch using standard functions and nodes with the quantised versions of \textit{linear} and \textit{relu} activation layers imported from the brevitas training library.

\subsection{Design space exploration}
\begin{figure}[t!]
    \centering
    \vspace{-0.7cm}
    \includegraphics[scale = 0.6]{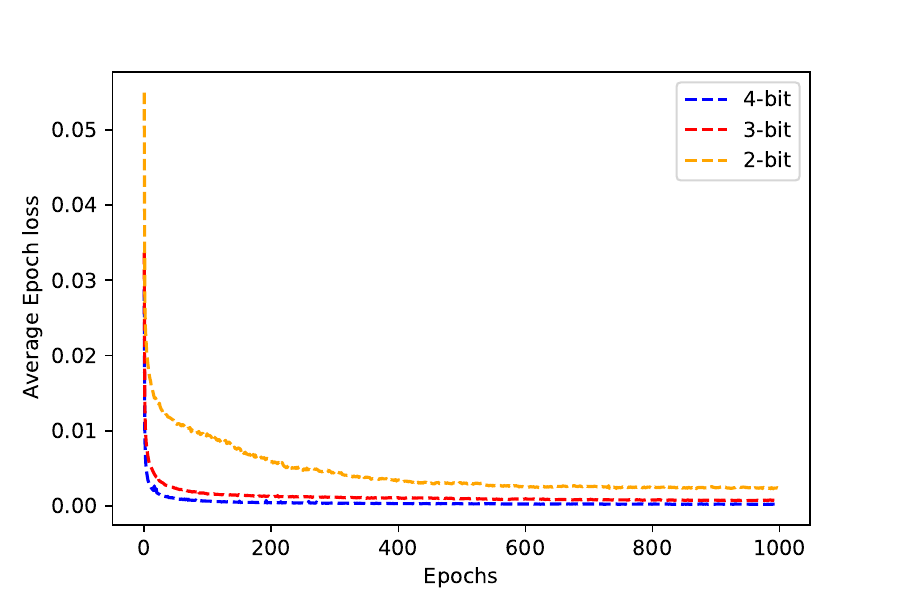}
    \caption{Training loss of the model for different precision of weights and activations for all the attacks.}
    \label{fig:train_loss}
\end{figure}

\begin{figure}[t!]
    \centering
        \vspace{-0.5cm}
    \includegraphics[scale = 0.6]{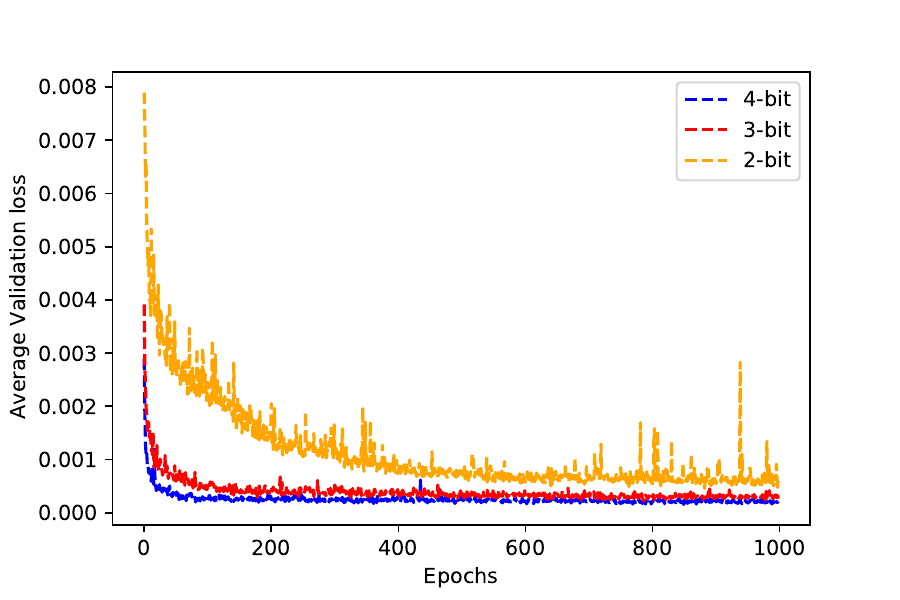}
    \caption{Validation loss of the model for different precision of weights and activations for all the attacks.}
    \label{fig:val_loss}
\end{figure}

To arrive at the optimum bit-precision for the quantised model, we trained our final MLP model at 2-bit, 3-bit, and 4-bit uniform quantisation on all the attacks to determine the trade-offs in detection accuracy and resource consumption (indirectly leading to power consumption and latency).
Figures~\ref{fig:train_loss} \&~\ref{fig:val_loss} show the training and validation losses when the model was trained for 1000 epochs at each quantisation. 
The 2-bit model converges to an optimum detection and is very close to the 3-bit \& 4-bit variants (\textit{between epoch 810 and 820}) in terms of validation loss. 
We observed \emph{147} misclassifications out of the total 180,000 messages in the test set for the 2-bit variant.
For the same test set, we found misclassifications to be \emph{74} \& \emph{31} for the 3-bit \& 4-bit variants respectively.

To capture the inference costs of the models further, we use the FINN utility function (inference cost) to estimate the memory footprint and operational complexity of each model.
This function captures the \textit{memory footprint} and \textit{binary operations} cost for each version of the model; the costs are then normalised (against a baseline) and linearly combined to arrive at the normalised inference cost for each of the three quantisation levels.
Table~\ref{table:inference cost} shows these costs with the 4-bit version as the baseline. 
We find the cost of the 2-bit version is 40\% \& 27\% of the 3-bit \& 4-bit version respectively, without a significant drop in detection accuracy and was hence chosen as the optimal bitwidth for deploying the CQMLP.

\begin{table}[t!]
\centering
\caption{Inference cost of the custom quantised models (CQMLP) inferred through the FINN library and normalised to the 4-bit model.}
\scalebox{1}{
\begin{tabular}{@{}lrrr@{}}
\toprule
\textbf{CQMLP Models} & \textbf{2-bit} & \textbf{3-bit} & \textbf{4-bit} \\
\midrule
Normalised inference cost  &    0.27      &     0.67      &    1\\
\bottomrule
\end{tabular}}
\label{table:inference cost}
\end{table}


\subsection{Dataset and Training}
\label{subsec:dataset}
We use the open Car Hacking dataset for training our model and to test its performance~\cite{canlink}. 
The dataset provides a labelled set of benign and attack messages which were captured via the Onboard Diagnostic (OBD) port in an actual vehicle, with attack messages injected in real-time. 
The dataset includes DoS, Fuzzing, and RPM-Spoofing message injections allowing us to validate the detection accuracy across these different attacks. 
We split the dataset as 85:10:5 for training, validation, and testing respectively, allocating the large section to training and optimisation of the quantised network. 
The performance of the model on the validation set during training ensures that it is not over-fitting on any of the attacks. 
We pre-process the dataset before training to mimic the dataflow the model will obtain as its input when integrated into the ECU. 
Each message (ID and payload) is encoded into INT8 type vector, and an adjacent sequence of \textit{n = 4} of these encoded messages form the input feature buffer content for the IDS (block shape of \textit{\{1,10*n\}} INT8 data).

The different bit models were trained using Brevitas, a quantisation-aware training (QAT) library from AMD/Xilinx\cite{brevitas}. 
We used the adam optimizer with binary cross-entropy as the loss function. 
The learning rate was set to 0.0001 to allow for slower learning which aids in efficient model training~\cite{wu2018training}.  
Each model has 54,824 learning parameters and was trained for 1000 epochs with a batch size of 128 for the DoS, Fuzzing, and RPM-Spoof attacks.
The best model in terms of validation loss was saved and exported as an \textit{'onnx'} graph.
This graph was then fed into the FINN hardware generation flow for the CQMLP-IDS IP generation.

\subsection{Dataflow hardware generation and integration to ECU ECU-IDS}
\begin{figure}[t!]
    \centering
    \includegraphics[scale = 0.55]{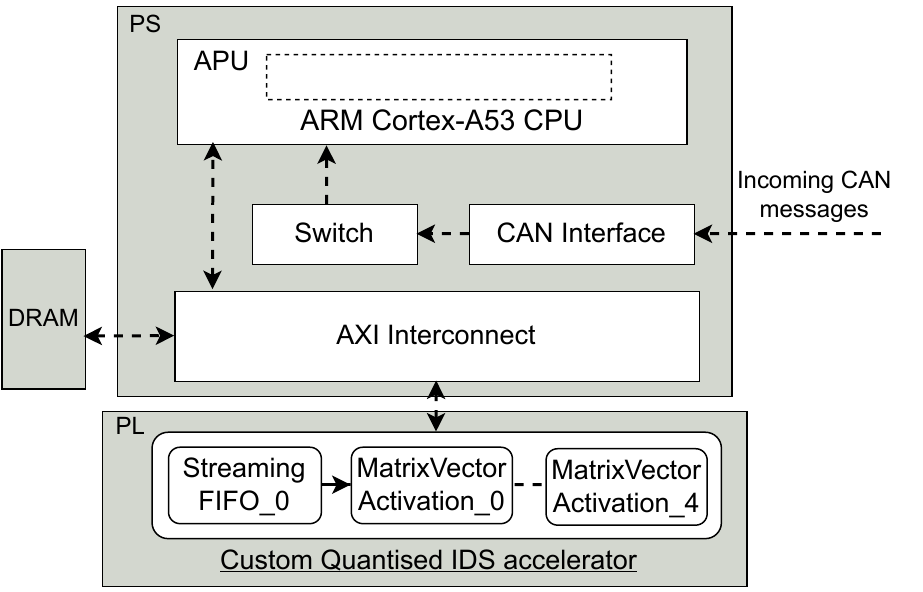}
    \caption{The proposed IDS-ECU architecture. The ML model is accelerated on the PL part of the FPGA.}
    \label{fig:arch}
\end{figure}

We utilise the FINN compiler to synthesize and generate the RTL description of the hardware implementation of the model.  
Starting at the \textit{'onnx'} graph, FINN applies a series of streamlining and dataflow transformations to convert the ML model into a synthesizable graph. 
We further specify the parallelism to achieve our target latency (set to 100000 messages per second), matrix-vector activation unit widths (set to 80), and folding levels 
to generate the dataflow hardware with AXI interfaces for input and output. 
Subsequently, FINN invokes Vivado tools to stitch the accelerator as an AXI slave to a Zynq subsystem for our target device (XCZU7EU) and generates the final bitstream along with the software drivers for our hardware ML engine.
This integration allows us to model the Zynq-based hybrid ECU architecture, where our custom hardware is accessible from the ECU and performs IDS in complete isolation from the tasks on the ARM cores as shown in Figure~\ref{fig:arch}.
We build on the proposed architecture from our previous works which contain detailed information of the same~\cite{khandelwal2022lightweight,khandelwal2023quantised}.
For our experiments, we use the PYNQ runtime on top of a Linux kernel on the ARM cores to interact with our hardware model on the PL.

\section{Experimental Results}\label{sec:experiments-date}
To quantify the performance and energy benefits of a dedicated dataflow accelerator implementing our CQMLP-IDS integrated with ECU function(s), we use a Zynq Ultrascale+ ZCU104 development board as the test ECU platform. 
The ZCU104 board features an XCZU7EV Ultrascale+ device with quad-core ARM A53 cores and dual-core ARM R5 cores on the PS as our target platform.
The A53 cores on the PS are configured to run at 1.2\,GHz peak, and the dataflow hardware was synthesized for a clock frequency of 200\,MHz by FINN.
The ARM cores run Linux OS with petalinux tools and PYNQ runtime to provide APIs to interface to the CQMLP-IDS IP in the PL. 
When testing, the hardware CQMLP model is fed with inputs message from the ECU mimicking the standard message flow within ECUs, and use our test split from the Car Hacking dataset to perform inference. 
We quantify the inference accuracy by evaluating the precision, recall, and F1 rates of our CQMLP models when inference is performed on the Zynq-IDS-ECU. 
We also quantify the per-message processing latency and power consumption for each incoming CAN message on the Zynq-IDS-ECU.
The results, in terms of inference accuracy and message processing latency, are compared against the state-of-the-art IDSs/IPSs presented in the research literature. 
We also compare the per-block processing latency and the per-inference energy consumption when the FP32 version of the proposed model is executed on an NVIDIA Jetson Nano 4GB GPU.
In the case of schemes where inference is performed on a block of CAN messages, we use these metrics along with the block size for the comparison.
We also compare our active power consumption against ML-based IDS approaches in literature where power consumption has been reported.

\subsection{Inference accuracy}
We evaluate the classification performance of the model in identifying \textit{DoS, Fuzzy} and \textit{RPM-Spoofing} attacks from the Car Hacking dataset and compare them against state-of-the-art techniques in the literature.
Table~\ref{table:confmatrix-2bit} captures the classification performance of our model in isolation across our test set as a confusion matrix. 
The misclassifications in our test set can be attributed to scenarios where nearly identical attack and benign message patterns are observed in a block at a given time which leads to an overall false positive rate (FPR) of 0.03\% (\emph{34 messages falsely classified as attacks out of the total 103176 benign messages}). 
We compare the inference performance of our 2-bit CQMLP model integrated within the ECU against the state-of-the-art IDSs and IPSs proposed in the literature: GIDS~\cite{seo2018gids}, DCNN~\cite{song2020vehicle}, MLIDS~\cite{desta2020mlids}, HyDL-IDS~\cite{lo2022hybrid} NovelADS~\cite{agrawal2022novelads}, TCAN-IDS~\cite{cheng2022tcan}, IForest~\cite{de2021efficient}, MTH-IDS~\cite{yang2021mth}, GRU~\cite{ma2022gru}, G-IDCS~\cite{10105249}, TAN-IDS~\cite{10141599} and Rec-CNN~\cite{desta2022rec} which are captured in table~\ref{table:comp}, comparing them in terms of inference precision, recall, F1 score and false negative rate.
For the \textit{DoS} attack the 2-bit CQMLP performs equally among the proposed IDSs in terms of F1 score.
In the case of \textit{fuzzing} attack, the 2-bit CQMLP model performs identically to DCNN~\cite{song2020vehicle} while performing better than iForest~\cite{de2021efficient}, TCAN-IDS~\cite{cheng2022tcan}, and GRU~\cite{ma2022gru} by 2.3\%, 0.6\% and 0.6\% respectively in terms of F1 score.
The 2-bit CQMLP also achieves almost perfect classification for the \textit{RPM-spoofing attack} and performs better than or equal to the other methods proposed in the literature.
Our prior work using two 8-bit variants of the feed-forward model achieved 99.96\%, 99.76\% \& 100\%  F1 scores for the DoS, Fuzzing, and RPM spoofing attacks, when deployed as two concurrent accelerators for detecting them simultaneously~\cite{khandelwal2022lightweight}. 
In contrast, our 2-bit CQMLP variant achieves almost identical F1 scores (99.90\%, 99.81\% \& 99.98\% respectively) for the same attacks (from the same dataset) while performing multi-class classification using a single inference model. 
By adopting QAT training and conservative block-based labelling, our low precision model is able to match the inference performance of the MLP variant at 8-bit quantisation discussed in~\cite{khandelwal2022lightweight}.

\begin{table}[t!]
\centering
\caption{Confusion matrix capturing the classification results of the CQMLP.}
    \scalebox{1}{
        \begin{tabular}{llcrr}
            \toprule
            \multicolumn{1}{c}{} & \multicolumn{4}{c}{\textbf{Predicted Values}} \\
            \cmidrule{2-5}   
            \textbf{CQMLP} & \textbf{Benign}  & \textbf{DoS}  & \textbf{Fuzzing} & \textbf{RPM-Spoof}  \\ 
            \midrule
            \textbf{Benign}   & 103142     &        17                          &  17   &          0                      \\ 
            \textbf{DoS}      &   27  &     23666                             &     0    &        0                \\ 
            \textbf{Fuzzing}  &  78   &        0                        &      28003    &        8                  \\
            \textbf{RPM-Spoof}  & 0    &              0                 &        0  &   25042                     \\
            \bottomrule
        \end{tabular}}
\label{table:confmatrix-2bit}
\end{table}

\begin{table}[t!]
\centering
\caption{Accuracy metric comparison (\%) of our CQMLP accelerators against the reported literature on the DoS, Fuzzing and RPM-Spoof attacks.}
\scalebox{1}{
\begin{tabular}{@{}llllll@{}}
\toprule
\textbf{Attack}  & \textbf{Model} & \textbf{Precision} & \textbf{Recall} & \textbf{F1}  & \textbf{FNR} \\
\midrule
\multirow{5}{*}{\textbf{DoS}} & GIDS~\cite{seo2018gids}                  & -                & 99.9          & -  &   -   \\ 
& DCNN~\cite{song2020vehicle}                  & 100                & 99.89          & 99.95  & 0.13     \\
& MLIDS~\cite{desta2020mlids}                  & 99.9                & 100          & 99.9  & -     \\
& G-IDCS~\cite{10105249}                  &      99.81          &     98.86      & 99.33  &  -    \\
& TAN-IDS~\cite{10141599}             & 100                & 100            & 100  & - \\
& HyDL-IDS~\cite{lo2022hybrid}                  & 100               & 100          & 100  & 0     \\
& NovelADS~\cite{agrawal2022novelads}                  & 99.97               & 99.91          & 99.94  & -     \\
& TCAN-IDS~\cite{cheng2022tcan}                  & 100             & 99.97          & 99.98  & -     \\
& iForest~\cite{de2021efficient}                  & -                &   -       &  - &  -    \\ 
& GRU~\cite{ma2022gru}                  & 99.93             & 99.91               & 99.92  & -     \\
& \textbf{CQMLP-IDS}                  &     99.92      &    99.88  &  99.90 &  0.11   \\
\midrule
\multirow{5}{*}{\textbf{Fuzzing}} & GIDS~\cite{seo2018gids}                  & -                & 98.7          & -   & -    \\ 
& DCNN~\cite{song2020vehicle}                 & 99.95             & 99.65          & 99.80  & 0.5     \\
& MLIDS~\cite{desta2020mlids}                  & 99.9             & 99.9          & 99.9  & -     \\
& G-IDCS~\cite{10105249}                  &       99.71         &     99      & 99.35  &   -   \\
& TAN-IDS~\cite{10141599}             & 99.99                & 99.99            & 99.99  & - \\
& HyDL-IDS~\cite{lo2022hybrid}                  & 99.98               & 99.88          & 99.93  &     \\
& NovelADS~\cite{agrawal2022novelads}                  & 99.99               & 100         & 100  & -     \\
& TCAN-IDS~\cite{cheng2022tcan}                  & 99.96             & 99.89          & 99.22  & -     \\
& iForest~\cite{de2021efficient}                  & 95.07                & 99.93          & 97.44  &    -  \\
& GRU~\cite{ma2022gru}                  & 99.32             & 99.13               & 99.22  & -     \\
& \textbf{CQMLP-IDS}            &       99.93      &   99.69      & 99.81  &    0.27  \\
\midrule
\multirow{5}{*}{\textbf{RPM-Spoof}} & GIDS~\cite{seo2018gids}                  & -                & 99.6          & -  &   -   \\ 
& DCNN~\cite{song2020vehicle}                  & 99.99                & 99.94          & 99.96  & 0.05     \\
& MLIDS~\cite{desta2020mlids}                  & 100                & 100          & 100  & -     \\
& G-IDCS~\cite{10105249}                  &     99.85           &     98.69      &  99.27 &  -    \\
& TAN-IDS~\cite{10141599}             & 99.99                & 99.93            & 99.96  & - \\
& HyDL-IDS~\cite{lo2022hybrid}                  & 100               & 100          & 100  & 0     \\
& NovelADS~\cite{agrawal2022novelads}                  & 99.9               & 99.9          & 99.9  & -     \\
& TCAN-IDS~\cite{cheng2022tcan}                  & 99.9             & 99.9          & 99.9 & -     \\
& iForest~\cite{de2021efficient}                  & 98.9                &   100       &  99.4 &  -    \\ 
& \textbf{CQMLP-IDS}                  &     99.96       &   100          & 99.98  & 0\\


\bottomrule
\end{tabular}}
\label{table:comp}
\end{table}

\subsection{Inference latency}
We quantify the per-message processing latency of the model, starting from the arrival of the CAN message at the interface to determine the detection delay incurred by the approach.
Table~\ref{table:latcomp} compares our result against other approaches in the literature, which utilise different platforms (GPUs, Jetson edge accelerators, Raspberry Pi) and approaches (block of CAN messages v/s individual messages). 
It should be noted that the latency metric in the case of block-based IDS does not consider the delay in acquiring the number of CAN messages required, and hence could result in potentially larger delays in detection of the onset of an attack. 
The tightly integrated CQMLP-IDS achieves 0.11\,ms per CAN frame, which is a 2.18$\times$ improvement over the dedicated line-rate QMLP-IDS ECU proposed in~\cite{khandelwal2022lightweight}.
We also observe the inference latency of the (fp32) model on the Jetson Nano (modeling a dedicated IDS ECU) to be 1.2\,ms when averaged over 10000 runs. 
This is 10.9$\times$ slower than the proposed CQMLP-IDS on the integrated IDS-ECU on the Zynq Ultrascale platform.
In terms of raw throughput, our dataflow implementation of CQMLP coupled to the ECU can process over 9090 messages per second at the highest payload capacity, achieving near-line-rate detection on high-speed critical CAN networks. 

\begin{table}[t!]
\centering
\caption{Per-message latency comparison against other state-of-the-art IDSs reported in literature.}
\scalebox{1}{
\begin{tabular}{@{}lrll@{}}
\toprule
\textbf{Models}     & \textbf{Latency} & \textbf{Frames} & \textbf{Platform} \\ \cmidrule{1-4}
GRU~\cite{ma2022gru} & 890\,ms & 5000 CAN frames & Jetson Xavier NX  \\
MLIDS~\cite{desta2020mlids} & 275\,ms & per CAN frame & GTX Titan X \\
Rec-CNN~\cite{desta2022rec} & 117\,ms & 128 CAN frames & Jetson TX2 \\
NovelADS~\cite{agrawal2022novelads} & 128.7\,ms & 100 CAN frames & Jetson Nano \\
GIDS~\cite{seo2018gids} & 5.89\,ms & 64 CAN frames &  GTX 1080  \\
TAN-IDS~\cite{10141599} & 11.6\,ms & 128 CAN frames & - \\
DCNN~\cite{song2020vehicle} &  5\,ms & 29 CAN frames & Tesla K80 \\
TCAN-IDS~\cite{cheng2022tcan} & 3.4\,ms & 64 CAN frames & Jetson AGX \\
MTH-IDS~\cite{yang2021mth} &  0.574\,ms & per CAN frame & Raspberry Pi 3     \\
QMLP-IDS~\cite{khandelwal2022lightweight} &  0.24\,ms & per CAN frame & Zynq Ultrascale+     \\
\textbf{CQMLP-IDS} & 0.11 ms & per CAN frame & Zynq Ultrascale+    \\ 
\bottomrule
\end{tabular}}
\label{table:latcomp}
\vspace{-0.3cm}
\end{table}

\subsection{Power consumption, Resource Utilisation (PS/PL)}
We further quantify the active power consumption of our dataflow accelerator to determine the average energy consumption per inference and the hardware resources incurred by the dataflow design on the XCZU7EV device. 
We observe that our model consumed 2.15\,W when measured directly from the device's power rails (using the PYNQ-PMBus package) while performing inference and other tasks on the ECU (with Linux OS), thus consuming 0.23\,mJ of energy per inference.
This is a 3.9$\times$ improvement compared to QMLP-IDS which report a per-inference energy consumption of 0.9\,mJ.
In comparison, we also observed the per-inference energy consumption of our (fp32) CAE model on the Jetson Nano to be 2.76\,mJ when averaged over 10000 inference runs which is 12$\times$ more than the proposed CQMLP-IDS on the tightly integrated IDS-ECU on a hybrid FPGA device.
Among other reported results in the literature, our approach improves the active power consumption by 4.6$\times$ when compared to the GRU~\cite{ma2022gru} model on an Nvidia Jetson Xavier as a dedicated IDS node.
In terms of resources, the dataflow CQMLP model consumes less than 2\% of resources on the device, leaving behind enough resources to integrate dedicated accelerators for the ECU function. 

We also quantify the utilisation of the ARM cores for managing the dataflow CQMLP accelerator to estimate the overhead in consolidating the IDS capability on the ECU. 
It was observed that a single core utilisation peaked at 40\% when processing the completion interrupt from the IDS core (at the highest message rate modelling the worst case scenario), while other cores remained at IDLE ($\le 2\% $ utilisation). 
We believe that this could be further refined by processing the classification results using hardware logic, interrupting the ECU only when a threat is detected, while also improving the detection latency.

\begin{table}[t!]
\centering
\caption{Resource utilisation breakdown of our proposed CQMLP (XCZU7EV).}
\scalebox{0.95}{
\begin{tabular}{@{}lrrrr@{}}
\toprule
\textbf{Node} & \textbf{LUT} & \textbf{FF} & \textbf{BRAM/URAM} & \textbf{DSP} \\
\midrule
2-bit-CQMLP-IDS  &    3999      &   4524         &      4/0      &  0 \\
\midrule
\% Usage & 1.74\% & 0.98\% & 1.28\%/ 0\% & 0\% \\
\bottomrule
\end{tabular}}
\label{table:resourceutilization}
\vspace{-0.3cm}
\end{table}


\section{Conclusion}\label{sec:conclusion}
In this paper, we present a dataflow-style hardware accelerator that implements a custom quantised MLP (CQMLP) model for detecting and classifying multiple attack vectors on an automotive CAN network. 
The accelerator generated through the FINN flow is quantised to 2-bit precision for weights and activations and is tightly integrated to the ARM core on a Zynq Ultrascale+ platform, mimicking a standard ECU. 
We show that the dataflow integration achieves a 2.2$\times$ speed-up in per-message processing latency and almost 3.9$\times$ reduction in energy consumption with 0.23mJ per inference when compared to the state-of-the-art IDSs. 
The quantised model also maintains high detection and classification accuracy across a range of attacks including DoS, fuzzing, and RPM-Spoofing attacks, making it an ideal approach for distributing IDS capabilities across ECUs in a vehicle network. 

\bibliography{references}
\bibliographystyle{ieeetr}

\end{document}